\RequirePackage{silence}
\documentclass[reprint,amsmath,amssymb,aps,pra,superscriptaddress,
    citeautoscript,
    floatfix]{revtex4-2}
\usepackage[utf8]{inputenc}
\usepackage[english]{babel}
\usepackage{hyperref}
\usepackage{bm}
\usepackage{hyperref}
\usepackage{physics}
\usepackage{tabularx}
\usepackage[page]{appendix}
\usepackage{nicefrac}
\usepackage[per-mode=power, 
            exponent-product = \times, 
            inter-unit-product = \ensuremath{{\cdot}}, 
            tight-spacing = true,
            number-unit-product=~,
            parse-numbers=false,]{siunitx}
\AtBeginDocument{\RenewCommandCopy\qty\SI}
\ExplSyntaxOn
\msg_redirect_name:nnn { siunitx } { physics-pkg } { none }
\ExplSyntaxOff
\usepackage{multibib}
\usepackage[dvipsnames]{xcolor}
\usepackage{booktabs}
\usepackage{float}
\usepackage{graphicx}
\usepackage[capitalise]{cleveref}
\usepackage{mhchem}
\usepackage{tabularx}
\usepackage{ulem}
\usepackage{enumitem}
\usepackage[
    protrusion=true,
    activate={true,nocompatibility},
    final,
    tracking=true,
    kerning=true,
    spacing=true,
    factor=1000]{microtype}
\hypersetup{
 colorlinks=true, 
 linkcolor=blue, 
 citecolor=blue,        % color of links to bibliography
 filecolor=blue,      % color of file links
 urlcolor=blue}
\usepackage[raggedright]{titlesec}
\usepackage{caption}
\titleformat{\section}{\raggedright\bfseries\large}{\thesection}{0pt}{}[]
\titleformat{\subsection}{\raggedright\bfseries}{\thesection}{0pt}{}[]
\titlespacing*{\section}{0em}{0.75em}{0pt}
\titlespacing*{\subsection}{0em}{0.75em}{0pt}
\renewcommand{\textcite}[1]{\citenum{#1}}
\newcolumntype{Y}{>{\centering\arraybackslash}X}
\Crefformat{appendix}{Supplementary Information S.#2#1#3}

\begin{document}
\setcitestyle{comma,sort&compress}
%               _    _                  
%   __ _  _  _ | |_ | |_   ___  _ _  ___
%  / _` || || ||  _|| ' \ / _ \| '_|(_-<
%  \__,_| \_,_| \__||_||_|\___/|_|  /__/
                                      
\author{Shao-Chien~Ou}
\affiliation{Joint Quantum Institute, NIST/University of Maryland, College Park, USA}
\affiliation{Microsystems and Nanotechnology Division, National Institute of Standards and Technology, Gaithersburg, USA}
\author{Alin O. Antohe}%
\affiliation{American Institute of Manufacturing Integrated Photonics (AIM Photonics), Research Foundation SUNY, Albany, NY, USA}
\author{Lewis G. Carpenter}%
\affiliation{American Institute of Manufacturing Integrated Photonics (AIM Photonics), Research Foundation SUNY, Albany, NY, USA}
\author{Gr\'egory~Moille}%
\affiliation{Joint Quantum Institute, NIST/University of Maryland, College Park, USA}
\affiliation{Microsystems and Nanotechnology Division, National Institute of Standards and Technology, Gaithersburg, USA}
\author{Kartik~Srinivasan}%
\email{kartik.srinivasan@nist.gov}
\affiliation{Joint Quantum Institute, NIST/University of Maryland, College Park, USA}
\affiliation{Microsystems and Nanotechnology Division, National Institute of Standards and Technology, Gaithersburg, USA}
\date{\today}

%   _    _  _    _      
%  | |_ (_)| |_ | | ___ 
%  |  _|| ||  _|| |/ -_)
%   \__||_| \__||_|\___|
\title{300 mm Wafer-Scale SiN Platform for Broadband Soliton Microcombs Compatible with Alkali Atomic References}
% \title{Shared All-Optical Trap of Integrated Multi-Solitons for Metrology Applications}}
%         _         _                   _   
%   __ _ | |__  ___| |_  _ _  __ _  __ | |_ 
%  / _` || '_ \(_-<|  _|| '_|/ _` |/ _||  _|
%  \__,_||_.__//__/ \__||_|  \__,_|\__| \__|
\begin{abstract}
Chip-integrated optical frequency combs (OFCs) based on Kerr nonlinear resonators are of great significance given their scalability and wide range of applications. %
Broadband on-chip OFCs reaching visible wavelengths are especially valuable as they address atomic clock transitions that play an important role in position, navigation, and timing infrastructure. %
% While silicon nitride (SiN) device layers fabricated via low pressure chemical vapor deposition (LPCVD) have been \greg{widely used} for on-chip \greg{OFC} generation, up to \qty{200}{\mm} wafer platforms, scaling to larger wafer sizes and integration with electronic and active photonic devices remain challenging due to its high deposition temperature and high film stress. %
Silicon nitride (SiN) deposited via low pressure chemical vapor deposition (LPCVD) is the usual platform for the fabrication of chip-integrated OFCs, and such fabrication is now standard at wafer sizes up to \qty{200}{\mm}. %
However, the LPCVD high temperature and film stress poses challenges in scaling to larger wafers and integration with electronic and photonic devices. %
Here, we report the linear performance and broadband frequency comb generation from microring resonators fabricated on 300~mm wafers at AIM Photonics, using a lower temperature, lower stress plasma enhanced chemical vapor deposition process that is suitable for thick (\qty{\approx700}{\nm}) SiN films and compatible with electronic and photonic integration.
The platform exhibits consistent insertion loss, high intrinsic quality factor, %
% (\qty{\approx10^6}{} at wavelengths near \qty{1060}{\nm}),
and thickness variation of \qty{\pm2}{\percent} across the whole \qty{300}{\mm} wafer. %
% Displaying variation in device layer thickness on the order of \qty{30}{\nm} across the \qty{300}{\mm} wafer, 
We demonstrate broadband soliton microcomb generation with a lithographically tunable dispersion profile extending to wavelengths relevant to common alkali atom transitions. %
These results are a step towards mass-manufacturable devices that integrate OFCs with electronic and active photonic components, enabling advanced applications including optical clocks, LiDAR, and beyond.
\end{abstract}

\maketitle
\noindent Optical frequency combs (OFCs) play a crucial role in diverse fields owing to their evenly spaced spectral lines, broadband coverage, and tunability~\cite{DiddamsScience2020}. %
Chip-scale integration is crucial for real-world deployment, in creating compact, efficient, and scalable metrology systems such as optical atomic clocks~\cite{MoilleNature2023}, spectrometers~\cite{SuhScience2016}, among others~\cite{ObrzudNat.Photon.2019,RiemensbergerNature2020}. %
%astronomical instruments~\cite{ObrzudNat.Photon.2019}, and ultrafast distance measurements~\cite{RiemensbergerNature2020}.} %
A promising approach to realizing on-chip low-noise OFCs involves periodically extracting a dissipative Kerr soliton (DKS) circulating within a microring resonator~\cite{KippenbergScience2018}. %
Silicon nitride (SiN) has emerged as a material of choice for such microrings due to its combination of low linear loss~\cite{BoseLSA2024}, high refractive index~\cite{ye_low-loss_2019}, broad optical transparency~\cite{MunozSensors2017}, and strong Kerr effect~\cite{Ikeda_thermal_2008}. %
SiN can also be integrated into conventional microelectronic and photonic fabrication workflows, allowing direct incorporation with high-frequency electronic systems~\cite{Zhang_low-temperature_2024} and active photonic devices~\cite{Park_heterogeneous_2020}. %

There have been many demonstrations of SiN microcomb fabrication on \qty{100}{\mm} \cite{LiuNatCommun2021a}, \qty{150}{\mm}~\cite{ye_foundry_2023}, \qty{200}{\mm}~\cite{el_dirani_ultralow-loss_2019} platforms. %
However, significant challenges remain in scaling such results to a \qty{300}{\mm} foundry process, primarily due to the large film stress and high temperature associated with the growth of thick stoichiometric (Si$_3$N$_4$) films by low pressure chemical vapor deposition (LPCVD)~\cite{ji_ultralowloss_2023}, the standard approach for SiN DKS microrings. %
Despite recent progress with \qty{300}{\mm} LPCVD-fabricated chips~\cite{liu_implementing_2025,Ou_CLEO_2025}, improvements in yield and dispersion engineering are needed to take full advantage of the \qty{300}{\mm} scale fabrication. %
One potential approach is to investigate modified LPCVD deposition conditions, which have been studied in both the stoichiometric~\cite{moille2021impactOL} and non-stoichiometric regimes~\cite{ye_low-loss_2019}, though a comprehensive process remains elusive. %
On the other hand, plasma enhanced chemical vapor deposition (PECVD)~\cite{chiles_deuterated_2018, XiePhoton.Res.PRJ2022, ji_ultralowloss_2023} and reactive sputtering~\cite{frigg_optical_2020, ZhangLaserPhotonicsRev.2024} present alternative solutions, but thus far research has primarily focused on telecommunications-band applications and has yet to expand to larger manufacturing-scale processes.
% have been explored as potential solutions, \ks{but thus far} the works have been mainly focused on telecommunications-band applications and are yet to be implemented in a \qty{300}{\mm} process.

\begin{figure}[t!]
    \centering
    \includegraphics{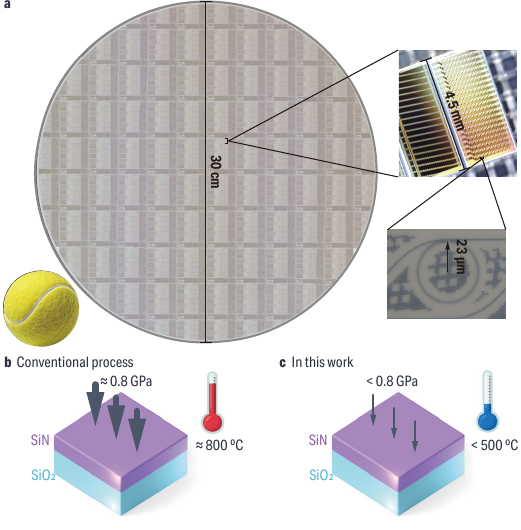}
    \caption{\label{fig:1}%
    Near-infrared PECVD SiN microcomb fabrication across a \qty{300}{\mm} silicon wafer. \textbf{a} Photographs of the \qty{300}{\mm} wafer alongside a tennis ball for relative scale (center and left) and optical microscope images of an individual die (top right) and one of the microring resonators (bottom right) studied in this work. \textbf{b} The conventional LPCVD process for SiN growth involves high tensile stress and growth temperature. \textbf{c} In this work, we employ a PECVD process with reduced film stress and lower deposition temperature.
    }
\end{figure}

\indent In this Letter, we demonstrate broadband DKS microcomb generation in a thick (\qty{\approx700}{\nm}) \ce{SiN} platform based on \qty{300}{\mm} wafer-scale fabrication at AIM Photonics~[\cref{fig:1}a]. %
The \ce{SiN} film is grown on a \qty{5}{\um} \ce{SiO2} layer produced by thermal oxidation of a \qty{300}{\mm} silicon wafer, from which a total of sixty-four reticle fields are harvested. %
Each reticle field measures \qty{26}{\mm} $\times$ \qty{32}{\mm} and contains thirty-six \qty{2.25}{\mm} $\times$ \qty{4.5}{\mm} chips with microring geometries designed for dispersion enabling broadband OFCs that target common alkali atom transitions wavelengths, similar to prior demonstrations using a 100~mm LPCVD process~\cite{yu_tuning_2019,MoilleNature2023}. %
Here, utilizing a PECVD fabrication process~[\cref{fig:1}b, c] with reduced film stress and lower deposition temperature compared to the conventional LPCVD approach~\cite{Kar_advanced_2024} enables straightforward integration with additional materials commonly used in microcomb applications, such as III-V semiconductors for lasers \cite{beeck_heterogeneous_2020} and detectors \cite{cuyvers_heterogeneous_2022}. %
We perform wafer-scale linear characterization and observe consistent low insertion loss and propagation loss suitable for Kerr comb applications. %
Pumped at \qty{283}{\THz} (\qty{1060}{\nm}), we generate soliton Kerr combs that extend to common alkali atom transition wavelengths, including Cs D2 (\qty{852}{\nm}), Cs D1 (\qty{894}{\nm}), and Rb D1 (\qty{795}{\nm}), marking a milestone toward mass production of thick SiN nonlinear devices for timekeeping, spectroscopy, and quantum applications.

%  A single chip measures approximately \qty{2.25}{\mm} $\times$ \qty{4.5}{\mm} and contains multiple devices. The devices we studied in this work have geometries designed to generate broadband frequency combs [\cref{fig:1}a] similar to those reported in \cite{MoilleNature2023}. The silicon nitride film is fabricated with significantly reduced film stress ($<$ \qty{0.8}{\GPa}) and lower deposition temperature ($<$ \qty{500}{\celsius}) [\cref{fig:1}b].

\begin{figure}[b!]
    \centering
    \includegraphics{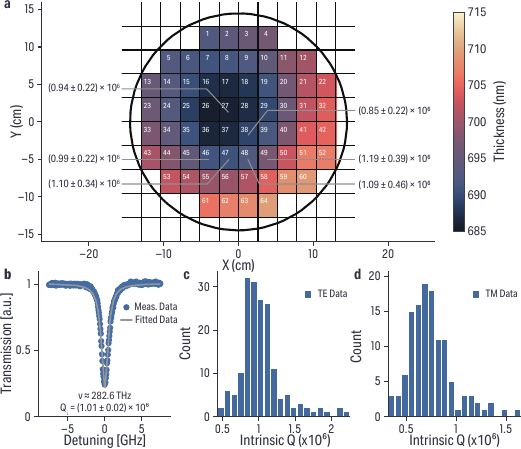}
    \caption{\label{fig:2}%
    Linear performance of the 300~mm silicon nitride platform. \textbf{a} Wafer map showing the device layer thickness and the measured mean and one standard deviation resonator $Q_\mathrm{i}$ in the pump band near \qty{283}{\THz}, for the TE$_{0}$ mode across each corresponding reticle field. \textbf{b} Measured and fit representative TE$_{0}$ resonance at \qty{282.6}{\THz}, yielding $Q_\mathrm{i}=(1.01\pm0.02) \times 10^{6}$, where the uncertainty is the \qty{95}{\percent} confidence interval of a least squares fit of the data. \textbf{c} Histogram of $Q_\mathrm{i}$ values in the pump band from 173 TE$_{0}$ resonances across the wafer. \textbf{d} Histogram of $Q_\mathrm{i}$ values in the pump band from 121 TM$_{0}$ resonances across the wafer.
    }
\end{figure}

\indent We first characterize the linear performance of the devices through measurements of fiber-to-chip facet coupling insertion loss and microring intrinsic quality factor ($Q_\mathrm{i}$). The nominal device geometry has ring radius \qty{RR=23}{\um}, ring width \qty{RW=820}{\nm}, gap \qty{G=350}{\nm}, and a bus waveguide of width \qty{W=500}{\nm}, with these parameters optimized for subsequent microcomb generation. The SiN film exhibits $\approx\pm~$\qty{2}{\%} variation around the nominal \qty{700}{\nm} target thickness [\cref{fig:2}a], which is comparable to results from PECVD films across 200~mm wafers and somewhat larger than the best values demonstrated for LPCVD films~\cite{ferraro_imec_2023}.  Devices from different reticle fields and with varying device layer thickness were measured to validate performance consistency across the wafer. The insertion loss is \SI{(2.02 \pm 0.11)}{\dB} per facet when measured using lensed fibers with a \qty{2.5}{\um} focused waist diameter, where the uncertainty is a one standard deviation value from measurements across 20 devices. The devices have a nominal \qty{W=200}{\nm} minimum width inverse taper waveguide edge coupler, which is designed to minimize insertion losses despite the non-square waveguide cross section. %
% The insertion loss was measured to be \qty{4.04}{\dB} $\pm$ \qty{0.21}{\dB} when coupling the pump signal in and out of the waveguide using lensed fibers. 
% The mean and standard deviation of the intrinsic Q in the pump band near \qty{283}{\THz} for fundamental transverse electric (TE$_{0}$) resonances corresponding to each reticle are shown in [\cref{fig:2}a]. 
$Q_\mathrm{i}$ is extracted by fitting each measured fundamental transverse electric (TE$_{0}$) and fundamental transverse magnetic (TM$_{0}$) resonance over the wavelength range of \qtyrange{1020}{1070}{\nm}, as illustrated in Fig.~\ref{fig:2}b. In total, 173 TE$_{0}$ resonances and 121  TM$_{0}$ resonances were measured across multiple devices and chips. %
% For example, a TE$_{0}$ resonance at a center frequency of \qty{282.58611}{\THz} is fitted, yielding an intrinsic Q of (1.01 $\pm$ 0.02) $\times$ 10$^{6}$. 
 We observe that average $Q_\mathrm{i}$ across the wafer remains consistently near $1\times10^6$, agnostic of the SiN thickness variation across the wafer [\cref{fig:2}a]. The TE$_{0}$ resonances exhibit $Q_\mathrm{i}$ values centered around 1 $\times$ 10$^{6}$, with the most probable value of 0.84 $\times$ 10$^{6}$ [\cref{fig:2}c]. For TM$_{0}$ resonances, $Q_\mathrm{i}$ values cluster around 0.65 $\times$ 10$^{6}$, with the most probable value of 0.69 $\times$ 10$^{6}$ [\cref{fig:2}d]. These results are comparable to the $Q_\mathrm{i}$ reported for devices with the same nominal geometric dimensions but with an LPCVD Si$_3$N$_4$ device layer~\cite{yu_tuning_2019,MoilleNature2023}.
 
 %SiN devices used for microcomb generation in the telecom band \cite{XiePhoton.Res.PRJ2022}. 
\begin{figure}[t!]
    \centering
    \includegraphics{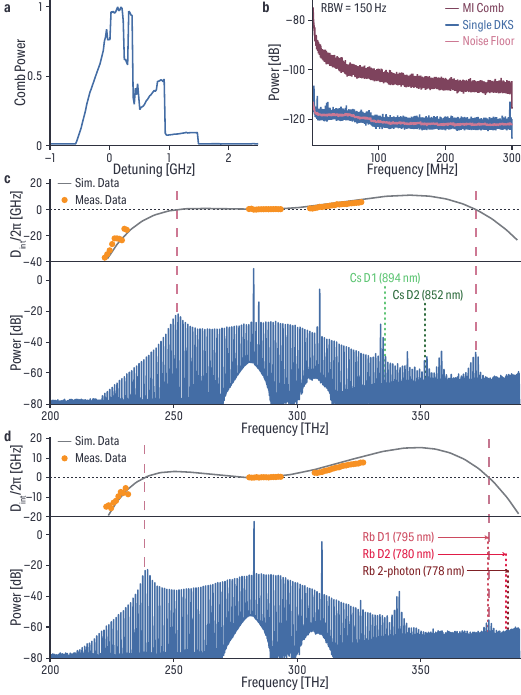}
    \caption{\label{fig:3}%
    Characterization of single DKS comb generation in the 300~mm silicon nitride platform. \textbf{a} Measured comb power evolution from the blue- to red-detuned regime of the pump laser with respect to the pump mode. %
    \textbf{b} Broadband noise spectra of the MI state, the single DKS state, and the measurement noise floor. The \qty{0}{\decibel} level corresponds to \qty{1}{\mW}, i.e., dBm. %
    \textbf{c}-\textbf{d} Optical spectra of single DKS combs (top) and the corresponding measured and simulated \(D_{\text{int}}\) (bottom) for \textbf{c} reticle field 46 and \textbf{d} reticle field 50. Orange points (solid lines) are measurements (simulations). Common alkali atom transition frequencies are highlighted in each spectrum. We note the additional creation of a synthetic DW arising from the phase matching of the off DKS-frequency grid cooler pump~\cite{MoilleNat.Commun.2021,MoilleNat.Photon.2025}.} 
\end{figure}

\indent The microring geometry described above was selected due to its weak anomalous group velocity dispersion near \qty{283}{\THz}, with higher-order dispersion enabling dispersive wave phase-matching at both ends of the spectrum. This design thus targets broadband soliton microcomb generation, with a repetition rate of \qty{\approx 1}{THz} determined by the selected $RR$. To study the devices experimentally, they are pumped with \qty{\approx 250}{mW} on-chip power. The comb power evolution is obtained by continuously scanning the pump laser from the blue- to red-detuned side of the pump mode near \qty{283}{\THz} [\cref{fig:3}a], while using a counter-propagating and cross-polarized \qty{310}{\THz} cooler laser for thermal-stabilization~\cite{zhang_sub-milliwatt-level_2019,zhou_soliton_2019}. Multiple steps are observed, indicating the generation and transition between modulation instability (MI) states and various soliton states, with the lowest step being the single DKS regime. We characterize the broadband noise of both a MI comb and a single DKS comb through direct photodection of the whole comb with the pump filtered out. As shown in Fig.~\ref{fig:3}b, the MI comb exhibits a high power-spectral density of white noise compared to the single DKS comb, with a \qty{\approx 20}{\dB} difference only limited by our  detection noise floor, highlighting the low-noise nature of the single DKS state.
%  The single DKS comb exhibits intensity noise at the same level as the measurement noise floor , confirming that the noise level of the single soliton state is below the detection limit of our experimental setup. This validates the generation of single DKS comb in the system. 

\indent We recorded the optical spectra of single DKS combs generated from identical microring resonator designs, with two representative examples shown in Fig.~\ref{fig:3}c,d. The combs, generated on reticle fields 46 and 50, have the same in-plane geometric parameters but a thickness difference of \qty{\approx10}{\nm}. Each exhibits broadband spectral coverage, with the widest spans going from approximately \qty{210}{\THz} to \qty{380}{\THz}. Notably, the generated combs overlap with common alkali atom transitions including the Cs D1 and D2 and Rb D1 lines used in atomic clock and quantum sensing applications. We also extracted the integrated dispersion $ D_{\mathrm{int}}(\mu)=\omega_{res}(\mu)-(\omega_{0}+D_{1}\mu)$ of the devices through wavemeter calibrated measurements, and we compare them against the simulated designs~[\cref{fig:3}c,d]. Here, $\mu$ is the mode index relative to the pump mode $\mu=0$, $\omega_{res}$ are the mode resonance frequencies, $\omega_{0}$ is the pump mode resonance frequency, and $D_1$ is the resonator free-spectral range extracted around the pump. The measured $D_{\mathrm{int}}$ values align well with the simulated dispersion profiles that account for the thickness difference between reticle fields. We also identified the zero-crossing points of each $D_{\mathrm{int}}$ curve and compared them with the dispersive wave locations observed in the comb spectra, confirming the consistency between measurement and simulation. 

\begin{figure*}[t!]
    \centering
    \includegraphics{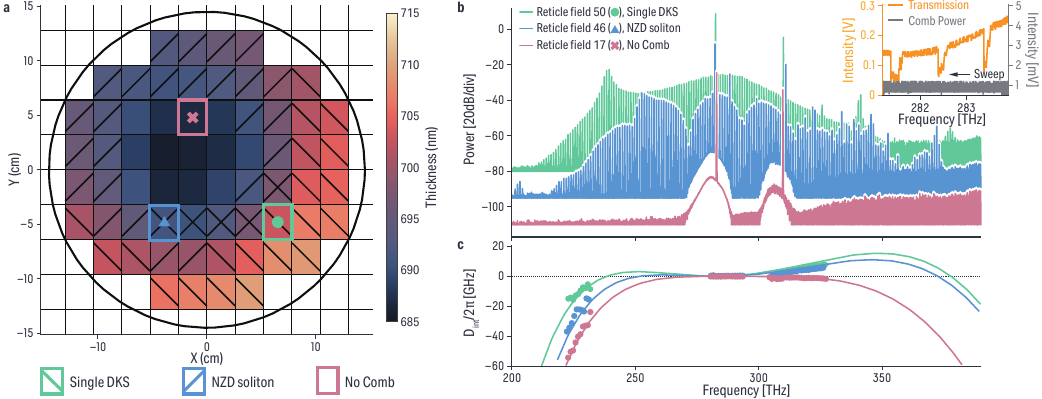}
    \caption{\label{fig:4}%
    Wafer thickness variation and corresponding nonlinear and dispersion characteristics. \textbf{a} Wafer thickness map and nonlinear state accessed for each reticle field. \textbf{b} Optical spectra of devices from three representative reticle fields (blue circle, pink triangle, orange cross), pumped near \qty{283}{\THz}. The inset displays the comb power evolution (gray) scanning from the blue- to red-detuned regime alongside the pump mode resonance spectrum (orange) in the case of no nonlinear state. \textbf{c} Measured (dots) and simulated (lines) \(D_{\text{int}}\) of the three selected devices pumped near \qty{283}{\THz}.
    }
\end{figure*}

\indent We next study the impact of the SiN thickness variation across all reticle fields. The accessible microcomb states and their resulting spectra depend critically on the resonator dispersion. As we choose a fixed nominal $RW$ and $RR$ and assume fixed material dispersion, thickness variation [\cref{fig:2}a] should be the primary cause of varying dispersion between reticle fields. By comparing the measured comb spectra and integrated dispersion (the latter reinforced by simulations accounting for the thickness variation) and how they evolve across the wafer, we categorized the nonlinear behavior into three regimes defined by device thickness: (1) single DKS generation, (2) interlocking of switching waves near zero-dispersion (NZD), and (3) no comb generation [\cref{fig:4}a].

%which \ks{for fixed material parameters,} is governed by the cross-sectional geometry of the ring waveguide~\cite{okawachi_octave-spanning_2011}. As mentioned previously, anomalous dispersion near the pump is essential for supporting bright solitons, where the interplay between Kerr nonlinearity and dispersion as well as parametric gain and loss enables stable pulse formation \cite{brasch_photonic_2016}. In contrast, normal dispersion disrupts this balance and, \ks{when proper excitation pathways are available, results in states such as dark solitons~\cite{xue_mode-locked_2015}.} In our work, the geometric dispersion is primarily controlled by RW and device layer thickness, given the rectangular cross-section of the waveguide. With fixed RW selected \ks{as mentioned above} and the device layer thickness varying across the wafer [\cref{fig:2}a], we \ks{studied} microcomb generation with devices from different reticles to investigate how the dispersion profile \ks{and resulting nonlinear state} evolves \ks{across the wafer}. By comparing the measured comb spectra across the wafer against the thickness map, we categorized the nonlinear behavior into three regimes defined by device thickness: (1) single DKS generation, (2) single DKS and collapsed snaking, and (3) no comb generation [\cref{fig:4}a]. 

\indent Representative measured comb spectra and \(D_{\text{int}}\) from one device in each regime are shown in Fig.~\ref{fig:4}b-c, respectively. From reticle field 50, we obtained a broadband single DKS microcomb with the \(D_{\text{int}}\) curve similar to those shown above and those reported in Ref.~\cite{yu_tuning_2019}, further confirming bright DKS generation in this platform. From reticle field 46, we observed a substantially different microcomb spectrum  and \(D_{\text{int}}\) curve, resembling those reported in Ref.~\cite{anderson_zero_2022}. This spectrum is consistent with a nonlinear state formed by interlocked modulated switching waves in a near-zero dispersion regime, forming solitary states that exhibit a quantized number of peaks. Such near-zero dispersion around the pump~\cite{xiao_near-zero-dispersion_2023} is explicitly corroborated by the \(D_{\text{int}}\) measurement. From reticle field 17, no comb spectrum was observed, with the \(D_{\text{int}}\) curve indicating the pump is situated in a normal dispersion region, consistent with the inability to support comb generation based on our excitation mechanism. This is further validated by the absence of nonlinear conversion in the recorded comb power evolution alongside the pump mode resonance spectrum in the inset of Fig.~\ref{fig:4}b. This transition from anomalous to normal dispersion with decreasing device thickness is consistent with other studies, including Refs.~\cite{okawachi_octave-spanning_2011,brasch_photonic_2016,moille_tailoring_2021}.

Figure~\ref{fig:4}a summarizes the accessible nonlinear states across the wafer. Single bright DKSs are observed on 26 of the 64 reticle fields, near-zero dispersion states are observed on 30 of the 64 reticle fields, and no nonlinear state is observed on 13 of the 64 reticle fields. Notably, both single bright DKSs and near-zero dispersion states are observed on 5 of the 64 reticle fields. As other parameters, such as $RW$, also impact the dispersion, introducing sufficient $RW$ variation can enable access to specific states of interest across a greater fraction of the wafer. Moreover, by biasing the average SiN thickness (e.g., towards a thicker film), desirable states such as single DKSs can potentially be accessed across the entirety of a 300~mm wafer. Of course, applications that rely on precise higher-order dispersion, e.g., to produce dispersive waves at specific frequencies, will require improvements to thickness uniformity across the wafer.

\indent In conclusion, we demonstrated broadband Kerr comb generation in a \qty{300}{\mm} foundry-based SiN platform. The fabrication process employed reduced film stress and lower deposition temperature, enabling future integration with other materials for advanced applications. Despite thickness variations among different reticle fields, we observed the generation of broadband single DKS combs across an appreciable fraction of the wafer. Notably, the single DKS comb exhibited a short dispersive wave near common alkali atom transitions. The measured thickness variation is similar to that reported in other well-established fabrication processes, and also provides valuable flexibility for tailoring dispersion profiles, enabling access to diverse soliton states and tunability of comb dispersive waves. 

%Across the wafer, the devices exhibited consistent insertion loss \greg{\qty{\approx2.02}{\dB}} and intrinsic  $Q_\mathrm{i} \approx 10^{6}$, comparable to other state-of-the-art systems.

\section*{Funding} National Institute of Standards and Technology (NIST-on-a-chip), Air Force Research Laboratory GDP4 and SDCP programs.

\section*{Acknowledgments} This material is based on research sponsored by Air Force Research Laboratory under AIM Photonics (agreement number FA8650-21-2-1000).

\section*{Disclosures} The authors declare no conflicts of interest.

% \bibliographystyle{apsrev4-2}
% \bibliography{Biblio}
%

\end{document}